\documentclass{article}

\usepackage{PRIMEarxiv}

\usepackage[utf8]{inputenc} 
\usepackage{soul}
\usepackage[T1]{fontenc}    
\usepackage{hyperref}       
\usepackage{url}            
\usepackage{booktabs}       
\usepackage{amsfonts}       
\usepackage{amsmath}        
\usepackage{nicefrac}       
\usepackage{microtype}      
\usepackage{lipsum}
\usepackage{fancyhdr}       
\usepackage{graphicx}       
\graphicspath{{media/}}     
\usepackage{multirow}
\usepackage{makecell} 
\usepackage{subfigure} 
\usepackage{xcolor}
\title{Dereverberation Filter by Deconvolution with Frequency Bin Specific Faded Impulse Response
}

\author{Stefan Ciba \\ \href{https://www.researchgate.net/profile/Stefan-Ciba}{https://www.researchgate.net/profile/Stefan-Ciba}
}

\begin{document}
\maketitle

\begin{abstract} 
This work introduces a robust single-channel inverse filter for dereverberation of non-ideal recordings, validated on real audio.
The developed method focuses on the calculation and modification of a discrete \textit{impulse response} in order to filter the characteristics from a known digital single channel recording setup and room characteristics such as early reflections and reverberations. The aim is a dryer and clearer signal reconstruction, which ideally would be the direct-path signal. The time domain \textit{impulse response} is calculated from the cepstral domain and faded by means of frequency bin specific exponential decay in the spectrum. The decay rates are obtained by using the blind estimates of reverberation time ratio between recorded output and test signals for each frequency bin. The modified \textit{impulse response} does filter a recorded audio-signal by deconvolution. The blind estimation is well known and stands out for its robustness to noise and non-idealities.
Application of the blind estimates for impulse response modification, in order to filter characteristics by deconvolution, offer advantages in scenarios, where adjustment for frequency bin specific reverberation and other characteristics is crucial, such as in audio production, acoustic simulation, or virtual reality applications. 
This filter accounts for not optimal recording conditions and can be applied as one of the first steps in processing of a single audio channel. Some real-time applications can benefit from this method, but some might need to have real-time or spatial adaption, when the \textit{impulse response} changes during the record and manual recalibration process would not be an option. 
Estimation of a direct path signal is key to many applications. 
\end{abstract}

\keywords{dereverberation \and spectral impulse response modification \and inverse filtering method}

\section{Introduction}
The field of digital audio signal processing has undergone significant development since the advent of digital techniques in the 1960s and 1970s, which enabled to manipulate sound with unprecedented precision and flexibility~\cite{oppenheim1989frequency,smith2011spectral}. 

Modeling and modification of the acoustic characteristics of room and recording setups, based on convolution with \textit{impulse response}, is a key technique in digital audio processing. The \textit{Linear Time Invariant} (LTI) system approach consisting of convolution of a audio signal with a measured or synthesized \textit{impulse response}, makes it possible to e.g. simulate reverberation, filter out early reflections, and compensate for coloration introduced by the recording chain~\cite{smith2011spectral,moorer1979reverberation}.

Traditional methods for reverberation and dereverberation often rely on parametric models or sparse representations of the \textit{impulse response}~\cite{moorer1979reverberation,HabetsThesis2007}. However, convolution with a sparse, spectrally modified \textit{impulse response}, that explicitly incorporates the $T60$ through exponential decay function offers a direct and flexible approach for single-channel dereverberation. This method enables targeted suppression of early reflections, reverberation, and undesired recording characteristics, based on established signal processing principles. This approach differs from closely related state of the art work such as "Speech Dereverberation in the Time-Frequency Domain" \cite{oyzerman2012}, "Approximation of Real Impulse Response Using IIR Structures" \cite{primavera2011} and "Representation and Identification of Systems in the Discrete-Time Wavelet Transform Domain" \cite{asm2007}.

\newpage
\section{Signal processing}
\subsection{Periodic sine sweep for system identification}
To characterize a system, a periodic linear sine sweep (chirp) can be used as the test signal \cite{farina2000}. 
The frequency of the positive chirp increases linearly from $f_0$ to $f_1$ within a duration $T$:
\begin{equation}
    \acute{x}(t) = \sin\left(2\pi \left[ f_0 t + \frac{f_1 - f_0}{2T} t^2 \right]\right), \quad t \in [0, T].
\end{equation}
The signals are being sampled with the sample rate $f_s=48$ kHz and thus contain $N=T f_s$ samples, counting $n=0,1,2, ... N-1$. By indexing with $t_n = \frac{n \bmod N}{f_s}$ the expression for the positive chirp becomes discrete and periodic:
\begin{equation}
\acute{x}_n = \sin\left(2\pi \left( f_0 t_n + \frac{f_1 - f_0}{2T} t_n^2 \right)\right).
\end{equation}
The chirp vector can be further extended for $P$ periods by array tiling:
\begin{equation}
x_{n} = \bigoplus_{p=0}^{P-1} \acute{x}_{n - pN}.
\end{equation}

\subsection{impulse response estimation}
Given discrete time series $x_n$ and its recorded signal $y_n$ with $N$ samples, their relation can be described by a discrete LTI system with $y_n=x_n*h_n$ \cite{Fliege1991}. Therein the \textit{impulse response} $h_n$ describes the signal chain, that can include unpleasant room characteristics that cause e.g. early reflections, reverberation, resonance and add non-linearity of the sensor and the loudspeaker.
\\
The \textit{impulse response} can be decorrelated by deconvolution from the related signals for system identification:  
\begin{equation}
h_n=y_n*x_n^{-1}.
\end{equation}
The deconvolution method of choice is the frame-wise subtraction of the test signal and the recorded signal in cepstral domain, followed by back transformation into time domain. The \textit{impulse responses} are subsequently combined and normalized to get one time domain representation. \\  
The discrete time signals are split into $N_{\text{frames}}$ frames with a frame length of $ N_{\text{DFT}}= 5 \cdot f_s $, so the DFT size was chosen in relation to the sample rate $f_s$, for the \textit{impulse response} to have a time duration that respects usual reverberation and delay time. 
For each frame index $ \eta = 0,1,2,..., N_{\text{frames}}-1 $ the frame boundaries are shifted over the signal with hop length $ N_{hop}=\frac{N_{\text{DFT}}}{2}$, which makes up a overlap of $o=50\%$. 
The frequency bins are counted by index $\mu=0,1,2,...,N_{\text{DFT}-1}$. In the same manner, the discrete time step is respected by the index $\nu$, to count the samples in a frame.
Subbands can be expressed as angular frequency $\Omega=\frac{2 \pi}{N_{\text{DFT}}}$ times index $\mu$.
DFT spectra are processed frame by frame with rectangular window $\omega^{\text{(rect)}}_\nu$. \\
The complex windowed short-time spectrum is:
\begin{equation}
\label{eq:STFT}
\underline{Y}_{\mu, \eta}=\sum_{\nu=0}^{N_{\text{DFT}}-1}y_{\eta N_{hop}-\nu} \cdot \omega^{\text{(rect)}}_\nu e^{-j\Omega \nu}.
\end{equation}

The original signal short time spectrum is $\underline{X}_{\mu, \eta}$.
The recorded short time spectrum is $\underline{Y}_{\mu, \eta}$. 
They are regularized by $|\underline{X}_{\mu, \eta}|_{\varepsilon} = \max(|\underline{X}_{\mu, \eta}|, \varepsilon)$ for numerical stability.

The real Cepstrum is obtained by transformation as follows: 
\begin{equation}
  C^{(x)}_{\mu,\eta} = \Re\left\{ \frac{1}{N_{\text{DFT}}}  \sum_{\mu=0}^{N_{\text{DFT}}-1} \ln|\underline{X}_{\mu, \eta}|_{\varepsilon} e^{j\Omega \mu} \right\}.
\end{equation}

%
%
%
%

In cepstral domain the deconvolutions are accomplished by subtractions, in order to obtain the \textit{impulse responses}:
\begin{equation}
  C^{(h)}_{\mu,\eta} =  C^{(y)}_{\mu,\eta} -  C^{(x)}_{\mu,\eta}.
\end{equation}
The \textit{impulse responses} transform back into the time domain vise versa. Although the imaginary part should be zero, only the real part is used, to prevent numerical errors:
\begin{equation}
h_{\nu,\eta} = \Re\left\{ \frac{1}{N_{\text{DFT}}} \sum_{\mu=0}^{N_{\text{DFT}}-1}  \overbrace{e^{\left(   \sum_{\mu=0}^{N_{\text{DFT}}-1}C^{(h)}_{\mu, \eta}  e^{-j\Omega \mu}    \right)}}^{H_{\mu, \eta}}   e^{j\Omega \mu} \right\}.
\end{equation}
The \textit{impulse responses} accumulate by averaging:
\begin{equation}
\overline{h}_{\nu}=\frac{1}{N_{\text{frames}}} \sum_{\eta=0}^{N_{\text{frames}}-1} h_{\nu,\eta}.
\end{equation}
To ensure the \textit{impulse response} has a consistent scale, with maximum absolute value $1$, it is normalized:
\begin{equation}
   h_{\nu} = \frac{\overline{h}_{\nu}}{\max_{\nu} |\overline{h}_{\nu}|}. \label{eq:IR}
\end{equation}
These steps yield a single, normalized \textit{impulse response}, that is suitable as a good overall representation.

\subsection{Spectral \textit{impulse response} modification}
The time domain \textit{impulse response}, e.g. from equation \eqref{eq:IR}, is transformed by STFT, to fade frequency bins based on blind $T60$ estimate.

\subsubsection{Blind estimation of $T60$ reverberation times for each bin}
The ISO 3382 standard \cite{iso3382} formally defines $T60$ as the time required for spatial sound energy to decay by 60 dB. 
This work uniquely uses blind estimate of $T60$, despite the \textit{impulse response} is known, to shape the \textit{impulse response} for each frequency bin, targeting specific acoustic and system artifacts for deconvolution filter, which is an approach not yet found in literature. Furthermore the $T60$ reverberation time is a quantity that is traditionally used for measurement, characterizing and shaping room acoustics usually as a global or band-averaged parameter ~\cite{kuttruff2016}.
Ratnam et al. propose an \textit{Maximum Likelihood} (ML) algorithm for a blind single-channel $T60$ estimate\cite{ratnam2003}.
To do the frequency bin specific modifications on the \textit{impulse response}, the necessary array of decay is obtained by the following procedure.
\\
The Power Spectral Density (PSD) for a discrete signal $x_n$ is its STFT magnitude squared:

\begin{equation}
S_{\mu, \eta} = |\underline{X}_{\mu, \eta}|^2 \label{eq:PSD}.
\end{equation}

The energy for each frequency bin is: 

\begin{equation}
E_{\mu, \eta} = \frac{\sum_{\eta' = \eta}^{N_{\text{frames}}-1} S_{\mu, \eta'}}{\sum_{\mu' = 0}^{N_{\text{DFT}}-1} S_{\mu, \eta'}}.
\end{equation}

The $T60$ time for each frequency bin is defined as the time it takes for the cumulative energy to fall below a threshold (e.g., 0.001, corresponding to -60 dB of magnitude):

\begin{equation}
\eta^{\text{(T60)}}_{\mu} = \min \left\{ \eta : E_{\mu, \eta} < 0.001 \right\}.
\end{equation}

The overlap factor is:
\begin{equation}
o = 1- \frac{N_{hop}}{N_{\text{DFT}}},
\end{equation}

the $T60$ estimate in seconds for every frequency bin is:

\begin{equation}
T60_{\mu} = \frac{o \cdot \eta^{\text{(T60)}}_{\mu} }{f_s}. \label{eq:T60mu}
\end{equation}

\subsubsection{Application of the decay function}
The ratio $\rho$ between the $T60$'s of the original test signal $x_n$ and its recorded signal $y_n$, weights the exponential decay for every frequency bin: $\rho^{\text{(T60's)}}_{\mu} =\frac{T60_{\mu}(y_n)}{T60_{ \mu}(x_n)}$.

With the frame wise time duration $ \tau = \eta \cdot \frac{N_{hop}}{f_s} $, the exponential decay function is:
\begin{equation}
D_{\mu, \eta} = e^{\left(-\frac{\tau}{\rho^{\text{(T60's)}}_{\mu}}\right)}.
\end{equation}

The frame wise decay modified \textit{impulse response} is:
\begin{equation}
h'_{ \nu , \eta} = \frac{1}{N_{\text{DFT}}} \sum_{\mu=0}^{N_{\text{DFT}}-1}  \underline{H}_{\mu, \eta} \cdot D_{\mu, \eta} \text{ } e^{-j\Omega \mu}.
\end{equation}
 
The synthesis window (Hann) is applied by $\omega^{\text{(hann)}}_\nu$ \cite{smith2011spectral} and furthermore the overlap has to be handled, done by \textit{Constant Overlap Add Method} (COLA):

\begin{equation}
h'_n = \sum_{\substack{\eta \\ 0 \leq n - \eta N_{hop} < N_{\text{DFT}}}}{\omega^{\text{(hann)}}_{n - \eta N_{hop}} \cdot h'_{n - \eta N_{hop}, \eta} }.
\end{equation}

\subsection{Filterbank application}
Additional overall exponential decay $dk$ can be applied to the modified \textit{impulse response} $h''_n= h'_n \cdot e^{-dk \cdot n}$ to avoid echoes. Finally, the modified \textit{impulse response} is applied as spectral filter bank $\underline{H}''_{\mu,\eta}$ for any recorded signal spectrum $\underline{Z}_{\mu,\eta}$ by deconvolution:
\begin{equation}
\underline{Z}''_{\mu,\eta}=\frac{\underline{Z}_{\mu,\eta}}{\underline{H}''_{\mu,\eta}}. \label{eq:system}
\end{equation}
Hann-windowing applies for STFT and iSTFT. The COLA method applies for iSTFT, respectively. The frame size was set to number of samples of the \textit{impulse response} and gain was normalized by strictly preventing clipping if necessary to get the desired filtered signal $\hat{z}''_{n} = z''_{n}  / max(|z''_{n}|)$.

\section{Test setup}

A sine-sweep and a speech audio was recorded by integrated microphone from a \textit{Convertible Workstation}
(\href{https://web.archive.org/web/20201031164451/https://h20195.www2.hp.com/v2/GetDocument.aspx?docname=c05939925}{HP ZBook Studio x360 G5 (6TW61EA\#ABD)}), on a desk, located in a pitched roof area corner of a living room, that has about 1 Ar. The sine-sweep sound was played back from the internal loudspeakers of the \textit{Convertible Workstation}. The integrated noise reduction was turned off. The goal was to improve the speech sound quality with the aim of having dry and clear sound when speaker sits in front of the \textit{Convertible Workstation}. A Futher assumption was, that degraded speech intelligibility could get better. \\ 

In the same manner the sine-sweep and a drum-set groove have been recorded from the same location. The sine-sweep was played back on a \href{https://ia601909.us.archive.org/31/items/manualsbase-id-372739/372739.pdf}{Alesis Active M1 MKII (8N)} near-field monitor beside the drum-set. The goal was to have dry and clear studio-like raw recording quality despite poor recording-conditions.

\section{Objectives}
Several objective measures prove the applicability of the Method.
Some Measures are used to compare the system impulse response with the filtered impulse response, and some are used to compare a filtered with a unfiltered signal. \\

A \textit{Logarithmic Signal-Power Attenuation} (LPA) was calculated to indicate the efficiency of the filter by taking the original signal $z_n$ and the filtered audio signal $z''_n$ into account:
	\begin{equation}
	LPA=10log_{10} \left(\frac{\overline{(z''_n-\overline{z''_n})^2}}{\overline{(z_n-\overline{z_n})^2}}\right).
	\end{equation}
The \textit{Short-Time Objective Intelligibility Measure} (STOI) measures the speech intelligibility \cite{stoi2010}. It is only applicable to speech audio.
The \textit{Perceptual Evaluation of Speech Quality} (PESQ) is an algorithm that models human auditory perception, capturing effects like distortion, background noise, clipping, and delay. It measures \textit{Mean Opinion Score, Listening Quality Objective} (MOS-LQO), which is basically the clarity of an audio channel, which is also necessary for non-speech audio channels. Higher MOS-LQO indicate better quality of the audio channel.
It is widely used in telecom, VoIP, and audio processing to measure speech enhancement or degradation \cite{pesq2001}. \\

Zwicker’s time-varying \textit{Loudness} model estimates how intense or loud a sound is perceived by human listeners \cite{ISO5322}. \\

\textit{Sharpness} from DIN 45692 standard (Zwicker’s model) quantifies how piercing or shrill sound feels, based on how present the frequencies above 3kHz are \cite{DIN45692Free}. \\

\textit{Roughness} measures how harsh or "fluttery" a sound feels, linked to rapid, audible amplitude fluctuations in the range most noticeable to human ears (20–300Hz modulation) \cite{Fastl2006}. \\

The \textit{Definition Ratio}, D50, is the ratio of early to total energy within 50 ms \cite{HoutgastSteeneken1985}. It was used to indicate how present reverberations and late reflexions are in the impulse response. \\

The Algorithms to measure perceptive sound quality can be used from the \textit{Modular Sound Quality Index Toolbox} \href{https://github.com/Eomys/MoSQITo}{(MoSQITo)} \cite{mosqito} \cite{mosqito2025}, \href{https://github.com/ludlows/python-pesq.git}{PESQ} \cite{pesq2025} and the \href{https://github.com/mpariente/pystoi}{STOI} \cite{stoi2025}, respectively. Further measures have been evaluated by proprietary software, namely the Cubase 5.0.1 build 147 (\href{https://o.steinberg.net/index.php?id=1782&L=1}{Steinberg Media Technologies GmbH}) \cite{cubase501}. \\

The measures to compare the unfiltered and filtered signals can be found in the Results section table \ref{tab:Objective}. 


\section{Results}

The filter has several well defined effects, which appear in the objective results table \ref{tab:Objective}.
Because reverberations are filtered from the signal, the signal amplitude and power decreases. 
The LPA measure shows that effect in one figure and gives it a name.
The most significant estimated pitch of the audio increases, which is due to filtered room characteristic that allowed deeper frequencies to fade slowly.
The filter decreases the $T60$ as expected, at least for those frequencies that showed reverberation.
Loudness, which can be unpleasant, decreases, but sharpness increases, as result of the filter.
Roughness is almost untouched by the filter, a slight decrease can be observed. 
The D50 shows, that reflections after 50ms disappeared.

\begin{table}[ht]
\centering
\caption{Objective measures. Statistics with asterisk (\textit{name}$^*$) derived from Cubase 5.0.1 build 147.}
\label{tab:Objective}
\begin{tabular}{|c|c|c|c|c|}
\hline
\multirow{2}{*}{ } & \multirow{2}{*}{\makecell{Drum Record \\ degraded ($z_n$)}} & \multirow{2}{*}{\makecell{Drum Record \\ filtered ($z''_n$)}} & \multirow{2}{*}{\makecell{Speech \\ degraded ($z_n$)}} & \multirow{2}{*}{\makecell{Speech \\ filtered ($z''_n$)}} \\
&       &       &       &       \\
\hline
\textit{min sample value}$^*$ &   -1    &   -0.919    &   -0.045   &    -0.045   \\
\hline
\textit{max sample value}$^*$ &   +1    &    +0.948    &    0.046   &     0.043   \\
\hline
\textit{peak amplitude}$^*$ &   0dB    &    -0.46dB   &   -26.75dB    &    -27.43dB     \\
\hline
\textit{DC offset}$^*$ &   -$\infty$dB    &    -$\infty$dB    &    -85.30dB   &    -86.44dB   \\
\hline
\textit{Estimated Pitch}$^*$ &    2248.0Hz / C\#6    &    9649.6Hz / D8   &    1445.2Hz / F\#5    &    1530.5Hz / G5   \\
\hline
\textit{Min RMS Power}$^*$ &   -$\infty$dB    &    -$\infty$dB    &    -61.03dB    &    -63.28dB   \\
\hline
\textit{Max RMS Power}$^*$ &    -1.84dB    &    -12.54dB   &    -32.82dB   &    -32.61dB   \\
\hline
\textit{Average}$^*$  &   -14.38dB    &    -28.84dB    &    -41.34dB   &    -41.27dB   \\
\hline
LPA &       \multicolumn{2}{c|}{-16.29dB}       &     \multicolumn{2}{c|}{ $7.63\cdot 10^{-5}$dB}      \\
\hline
$T60$  &    5.475ms   &   0.186ms    &    0.1816ms   &    0ms  \\
\hline
STOI &    N/A   &    N/A   &   0.9978    &    0.9978   \\
\hline
$\Delta$ STOI   &    \multicolumn{2}{c|}{N/A}       &      \multicolumn{2}{c|}{0}      \\
\hline
MOS-LQO &    4.4202   &    4.5314    &    4.6308    &    4.6337   \\
\hline
$\Delta$ MOS-LQO &      \multicolumn{2}{c|}{$0.1112$}      &      \multicolumn{2}{c|}{$2.9 \cdot 10^{-3}$}       \\
\hline
\textit{Loudness}  &    0.90320   &    0.30145   &    0.14764   &    0.14393   \\
\hline
\textit{Roughness} &    0.23022   &    0.22169   &    0.21995    &   0.21946    \\
\hline
\textit{Sharpness} &    3.26095   &    3.95266   &    3.09846   &    3.16039   \\
\hline
D50 							& 		$42.9\%$				&			$100.0\%$		&			$71.7\%$				&				$100.0\%$			\\
\hline
\end{tabular}
\end{table}

\begin{figure}[h]
\centering
\includegraphics[width=0.70\textwidth]{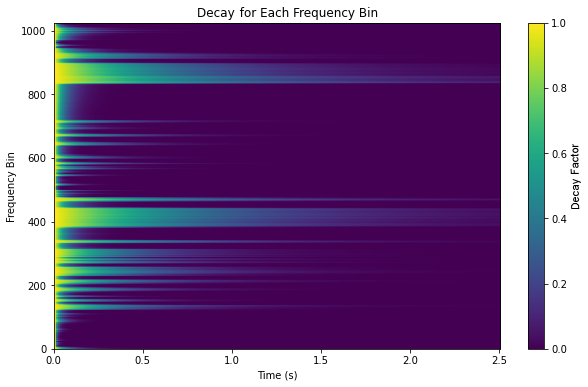}
	\caption{The $T60$ decay matrix.}
	\label{fig:Decaymatrix}
\end{figure}

\begin{figure}[h]
     \centering
     \subfigure[Original.]{\includegraphics[width=0.9\textwidth]{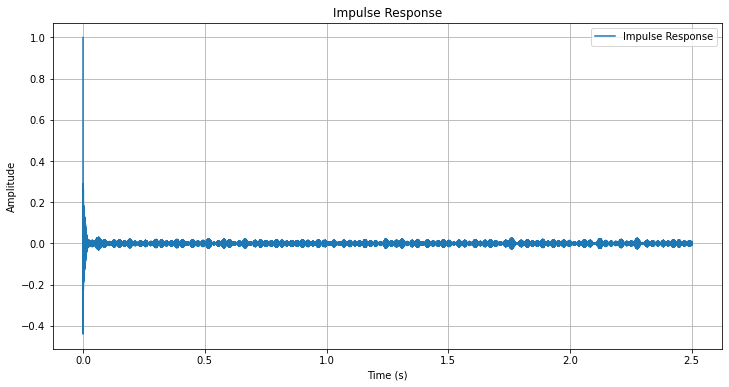}}
     \hfill
     \subfigure[Modified.]{\includegraphics[width=0.9\textwidth]{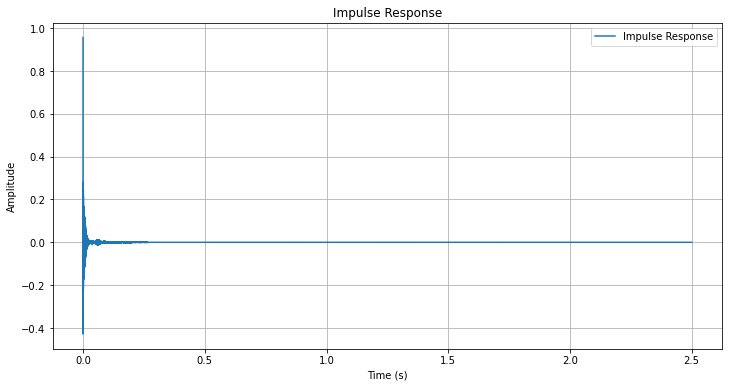}}
\\
\caption{\textit{Impulse response}.}
	\label{fig:results}
\end{figure}

The original and the modified impulse response of the speech audio recording setup are shown in numerical plot \ref{fig:results}.
Slow fading, early reflections and some unwanted behavior, that account for late reflections, can be seen in the original unmodified impulse response. 
The filter decreases the unwanted behaviors selectively for the frequency bins based on the $T60$ decay matrix in numerical plot \ref{fig:Decaymatrix}. 
Because the impulse response from the speech record was different from the drum record, filter results differ a lot.
The STOI measure stays untouched by the filter: No speech enhancement and no speech degradation due to filtering: No noise was added or subtracted to the speech sample due to filtering.
Unexpectedly, but for obvious reason, the $T60$ decreased to zero after filtering in the speech audio.

\section{Conclusion}

The filter is designed aggressively on purpose and filters out reverberation and produces a dry and clear sound with a minimum of room and recording setup characteristic, allowing the direct path signal to be dominant.
However, as the filter does exactly what it should, the output might need further processing for more pleasant sound characteristic, but is obviously a excellent start for further audio processing. 


\section{Supplements}
The Python-code of the studied method and audio samples are to be found at GitHub: \href{https://github.com/fanci90/Digital-Deconvolution-Audio-Filter-repo.git}{https://github.com/fanci90/Digital-Deconvolution-Audio-Filter-repo.git}.

\section*{Acknowledgments}
The author gratefully acknowledges Simon Ciba, M.A., known for e.g. "WhisPER – A New Tool for Performing Listening Tests", from the Audio Communication Group, TU Berlin, for his careful review of this work and his valuable comments.

\end{document}